\documentclass[dvips]{article}

\usepackage{icrc2011}
\usepackage{sidecap}

\title{VERITAS observation of Markarian 421 flaring activity}

\shorttitle{Galante, VERITAS observation of Mrk~421}

\authors{Nicola Galante$^{1}$ for the VERITAS Collaboration$^{2}$}
\afiliations{$^1$Harvard Smithsonian Center for Astrophysics\\
$^2$http://veritas.sao.arizona.edu}
\email{ngalante@cfa.harvard.edu}

\abstract{Markarian~421 is one of the brightest BL-Lac objects in $\gamma$-rays in the northern 
hemisphere. Because of its brightness, the source has been the focus of several
 coordinated multi-wavelength (MWL) campaigns designed to study 
the physical processes responsible for the non-thermal continuum emission. 
The blazar monitoring program of VERITAS recently received a ToO during a strong 
flaring event by Markarian~421 in February 2010. The source was seen at  flux level 
of approximately 8 Crab units and exhibited  spectral evolution and variability features. A 
multi-wavelength campaign with other MWL partners was udndertaken. 
Results on past and recent flaring events are presented.}

\begin{document}
\maketitle


\section{Introduction}

Blazars are a subclass of active galactic nuclei (AGNs) presenting rapid variability and non-thermal 
emission across nearly the entire electromagnetic spectrum, implying 
that the observed photons  originate within highly relativistic jets oriented very close to the 
observerÕs line of sight~\cite{UrryPadovani}. 
Various models have been proposed to account for the broadband spectral energy distributions (SEDs) 
observed in very high energy (VHE: $E>100$~GeV) blazars, which typically display two 
major components peaking at different energies: 
the lower-energy peak \mbox{($10^{13}$ Hz $\lesssim E_\mathrm{peak}\lesssim 10^{19}$ Hz)} 
is due to synchrotron emission by highly relativistic electron and positrons, 
and the higher-energy component is due to inverse-Compton (IC) 
scattering of the synchrotron photons by these relativistic $e^+e^-$~\cite{Jones1974,BloomMarscher1996}. 
Hadronic interactions producing neutral pions which decay into photons~\cite{Mannheim1992}, 
and synchrotron emission from protons~\cite{Aharonian2000} are also possible scenarios for the high-energy 
component of blazars. 

Observationally, blazars undergo both major outbursts on long 
time scales and rapid flares on short time scales, most prominently at keV and TeV energies.
During some outbursts, both of the SED peaks have been observed to shift toward higher energies 
in a generally correlated manner~\cite{Acciari2009}. The correlation of the variabilities at keV and TeV energies (or 
lack thereof) during such outbursts has aided in refining the emission models.
 In addition, rapid, sub-hour 
flaring activity is interesting as it provides direct constraints on the size of the emission region. 

BL-Lac objects are a particular class of blazars 
with the featureless non-thermal continuum dominating over the discrete emission.
Markarian~421 (Mrk~421; 1101+384), at a redshift of $z = 0.031$, is a high-frequency peaked
\mbox{(HBL: $E_\mathrm{peak} \gtrsim 10^{17}$ Hz)} BL-Lac object that historically shows
intense and rapid flaring episodes. Its flaring activity is particularly interesting for two  reasons: 
1) its relativistic high flux facilitates the characterization if its spectral and temporal features;
2) frequent flaring activity permits the time evolution of the low and high portions 
of the SED to be studied.
Therefore Markarian~421 has been the focus of coordinated MWL observational campaigns, 
from optical to the
$\gamma$-ray energy band. These campaigns are triggered by the observation of
flares by any of the MWL partners monitoring the source.

 Here we report results from past and recent observation of Markarian~421 flaring episodes.

\section{VERITAS Observation}

\subsection{The VERITAS Detector}

The VERITAS detector is an array of four 12-m diameter imaging
atmospheric-Cherenkov telescopes located in southern Arizona~\cite{Weekes}. 
Designed to detect emission from astrophysical
objects in the energy range from 100~GeV to greater than 30~TeV,
VERITAS has an energy resolution of $\sim$15\% and an angular
resolution (68\% containment) of $\sim$$0.1^\circ$ per event at 1~TeV.  A source
with a flux of 1\% of the Crab Nebula flux is detected in 25~hours
of observations.  The field of view of the VERITAS telescopes is
$3.5^\circ$.  For more details on the VERITAS instrument and the imaging atmospheric-Cherenkov
technique, see~\cite{Perkins2009}.

\subsection{Past observational campaigns of Markarian~421 flaring activity}

\begin{figure}[!t]
  \centering
  \includegraphics[width=3.0in]{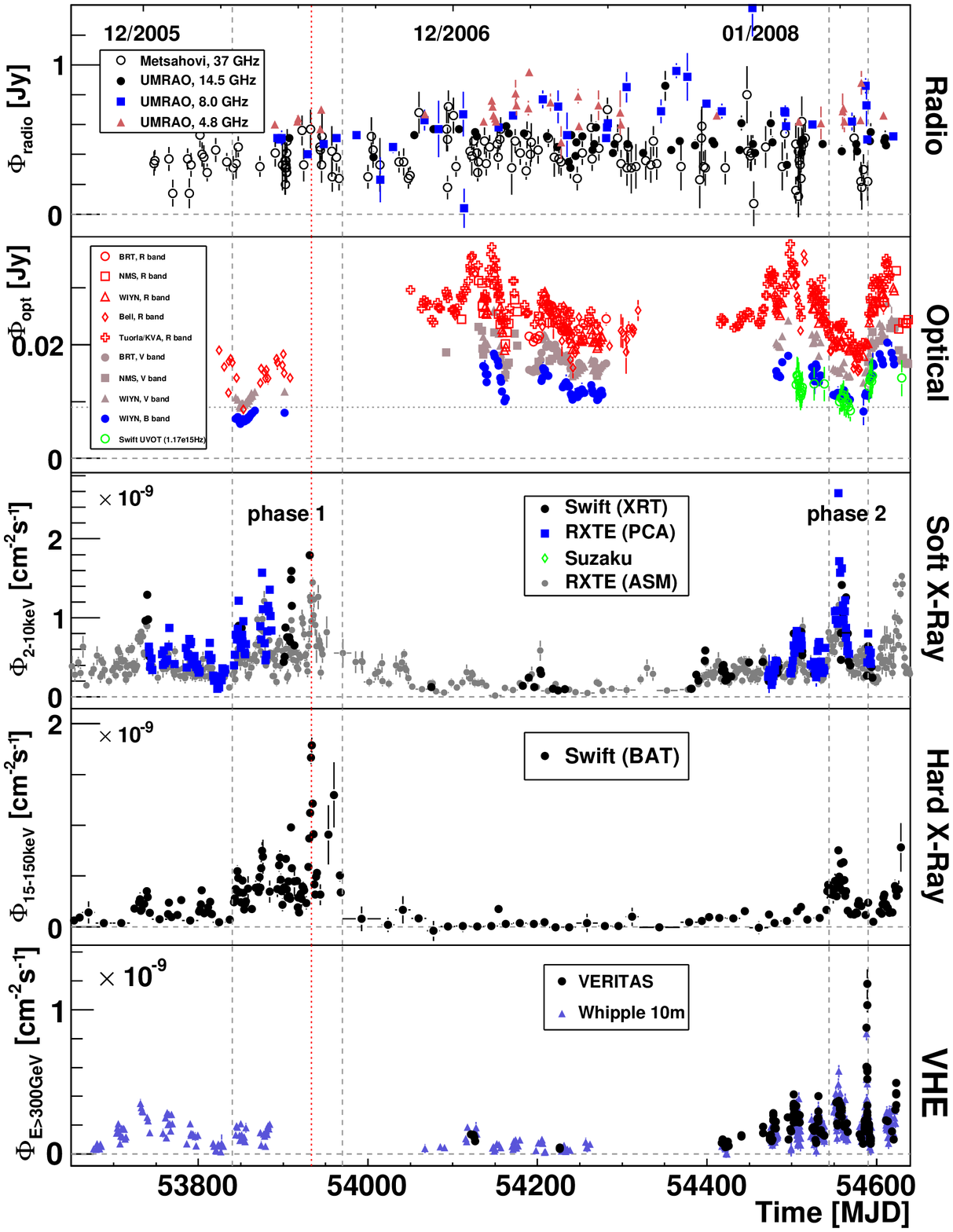}
  \caption{Light curves measured by different experiments in 2006-2008, including extensive VHE observations 
  by Whipple and VERITAS. Two phases of activity appear in the X-ray and TeV bands 
  (phases 1 and 2, grey vertical lines). Phase 1: Maximum of hard X-ray flux (Swift/BAT, red line). 
  Phase 2: Good X-ray/TeV coverage, X-ray observations partly triggered by VERITAS. Figure from~\cite{Matthias2011}}
  \label{fig:fig1}
 \end{figure}

Markarian~421 has been intensively studied in the past both by VERITAS and by the
Whipple 10~m $\gamma$-ray telescope. Between 2006 and 2008, Whipple and VERITAS
recorded 96~hrs and 47~hrs of data on Markarian~421 respectively. During this campaign,
quasi-simultaneus MWL data in radio, optical and X-ray have been taken. 
Figure~\ref{fig:fig1} shows the combined lighcurve of this long-term MWL campaign.
A detailed analysis of the MWL data
over this campaign is presented in~\cite{Matthias2011}. Flux variability is found
in all bands except in radio. In particular, the X-ray and VHE energy bands are found
to be often correlated. Such correlation implies that the particle population responsible for the
synchrotron and IC component are the same. Although this correlation is seen as a general trend, 
it does not necessarily hold true at the level of individual flares. On the other hand,
optical/TeV correlation is not found, suggesting that the optical emission might not be dominated by
the optical synchrotron component from the jet.
The broadband SED is well described by a single zone SSC model and no evidence for flux variability
on the time scale of a minute is found.

Our ToO program on Markarian~421 
was first triggered in April 2006 by a major outburst from Markarian~421 as detected by 
regular monitoring of the VHE band by the Whipple 10~m telescope. 
The program was triggered again in May 2008 by another major outburst from
 Markarian~421, also detected in the VHE band. 
Because of observational constraints, in both cases we captured only the decaying portion of the outburst. 
However, taken together, the two campaigns have produced a significant amount of simultaneous 
optical/UV, X-ray, and VHE data. During the first April 2006 campaign, 
Markarian~421 was observed for about 3~hrs with the Whipple 10~m $\gamma$-ray telescope, 
and for about 3~hrs with MAGIC, at energies above 400 and 100~GeV respectively. 
During the following May 2008 flare, about 2~hrs of observations by VERITAS 
were obtained. During each of the three VHE observations, truly simultaneous data in optical/UV 
were recorded with the \emph{XMM-Newton} satellite's Optical Monitor \mbox{(OM: 170-650 nm)}
and EPIC-pm detector \mbox{(EPN: 0.5-10 keV)}.

The coordinated ToO MWL campaign provided important, and to an extent unexpected, results 
on the spectral behavior of Markarian~421~\cite{Matthias}. The broadband SED is fitted with the one-zone
leptonic SSC model. The best fit to the MWL data set is obtained
under the assumption of a pure SSC mechanism without any extra external Compton (EC) component.
Surprisingly both 2006 and 2008 data sets show no obvious correlation between X-ray and VHE data,
nor between optical/UV and VHE. Clear correlation between optical/UV and X-ray data is nowhere observed.
This suggests that the X-ray and VHE photons originate from two distinct electron/positron polulations, 
similar to what was observed during a study of PKS2155-304~\cite{Aharonian2009}. Other scenarios that could explain 
the observed variability patterns include the possibility of an inhomogeneous emission 
region or hadronic origin of the VHE emission.
It is interesting to note that Markarian~421 during this particular flares
is clearly behaving very differently here from its typical flaring
periods (e.g. \cite{Acciari2009}), where the X-ray and VHE variabilities are 
seen to be strongly correlated. 
Spectral hysteresis patterns are also observed in the X-ray data during the 2006 flare. Although spectral 
hysteresis has been previously observed is other blazars too, the phenomenon is not fully understood
yet. A possible explanation is that spectral hysteresis is produced by the combined effect of three different
typical time scales~\cite{Kirk1999}: 
the duration of the flaring variability $t_\mathrm{var}$, the synchrotron cooling time
$t_\mathrm{cool}$ and the particle acceleration $t_\mathrm{acc}$. 
The clockwise hysteresis found in the X-ray data,
indicating a lag at lower energies in the X-ray band, coupled with the essentially symmetric shape of 
the flare in the X-ray light curve seems to indicate that the case with
 $t_\mathrm{cool}\gg t_\mathrm{var}\gg t_\mathrm{acc}$ is most relevant to this observation.

\subsection{February 2010 flaring episode}

VERITAS monitored the BL-Lac Markarian~421 from November 2009 to April 2010,
taking short snapshots at irregular intervals, for approximatively 20 hours of
good-quality effective observation time. The observations have been performed
by pointing the telescopes at $0.5^\circ$ North/South/East/West offset in order to get simultaneous background
measurement, with the telescopes operating at an average zenith angle of $17^\circ$. 
On February 17, 2010 VERITAS was alerted by its $\gamma$-ray partners that
the source was in flaring state. Five hours of data were obtained.
 Unfortunately, due to technical reasons
telescope \#1 was unavailable, resulting in a degraded 3-telescopes array observation.
The VERITAS flaring event triggered a MWL campaign on the following nights to its
X-ray partners~\cite{atel}.
In addition, the VERITAS observation was performed during an  X-ray monitoring
by the \emph{Swift} X-ray telescope (XRT). XRT recorded about 1~ks of quasi-simultaneous data
during the night of the flare. XRT took another 11~ks of data during the following two nights, when
Markarian~421 was still in flaring state although at a lower flux level. A total of 22~ks
of X-ray data were taken by XRT between February 15, 2010 and March 4, 2010.
Finally, the \emph{Fermi} large area telescope
(LAT) provided simultaneous observation of the source
in the \mbox{100 MeV - 100 GeV} energy range.

Prior to event selection and background subtraction, all shower images
are calibrated and cleaned as described in~\cite{Cogan:2006,Daniel:2007kx}.  
Several noise-reducing event-selection cuts are made at this point. Following
the calibration and cleaning of the data, the events are parametrized
using a moment analysis~\cite{Hillas:1985ta}.  From this moment
analysis, scaled parameters are calculated and used for the
selection of the $\gamma$-ray-like events~\cite{Aharonian:1997rm,Krawczynski:2006ts}. The event
selection cuts are looser than the ones typically used on a Crab-like source 
(power-law spectrum photon index $\alpha=2.5$ and 1~Crab Nebula flux level)
in order to take advantage of the large statistics in the flux reconstruction. 

\begin{figure}[!t]
  \vspace{5mm}
  \centering
  \includegraphics[width=3.in]{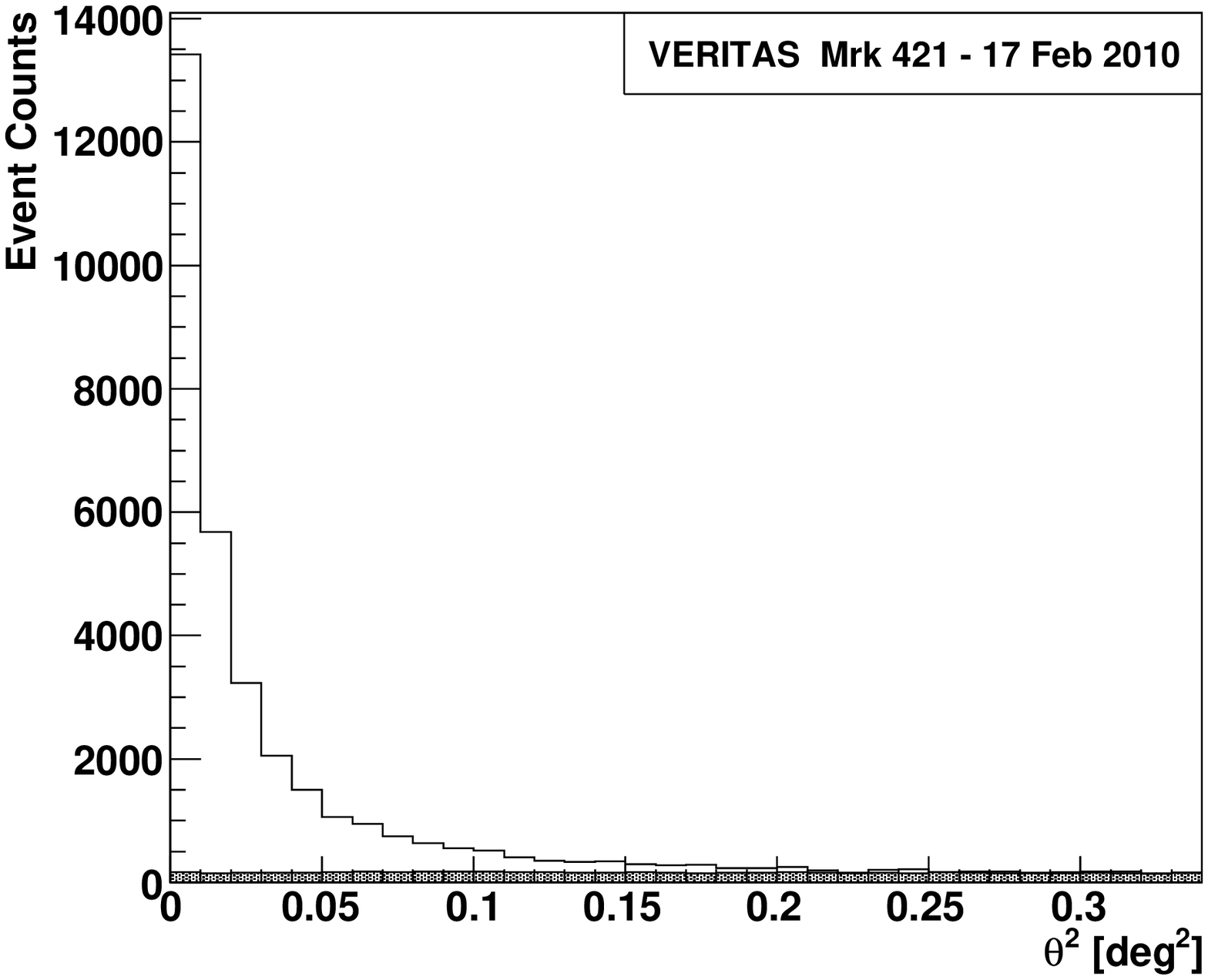}
  \includegraphics[width=3.0in]{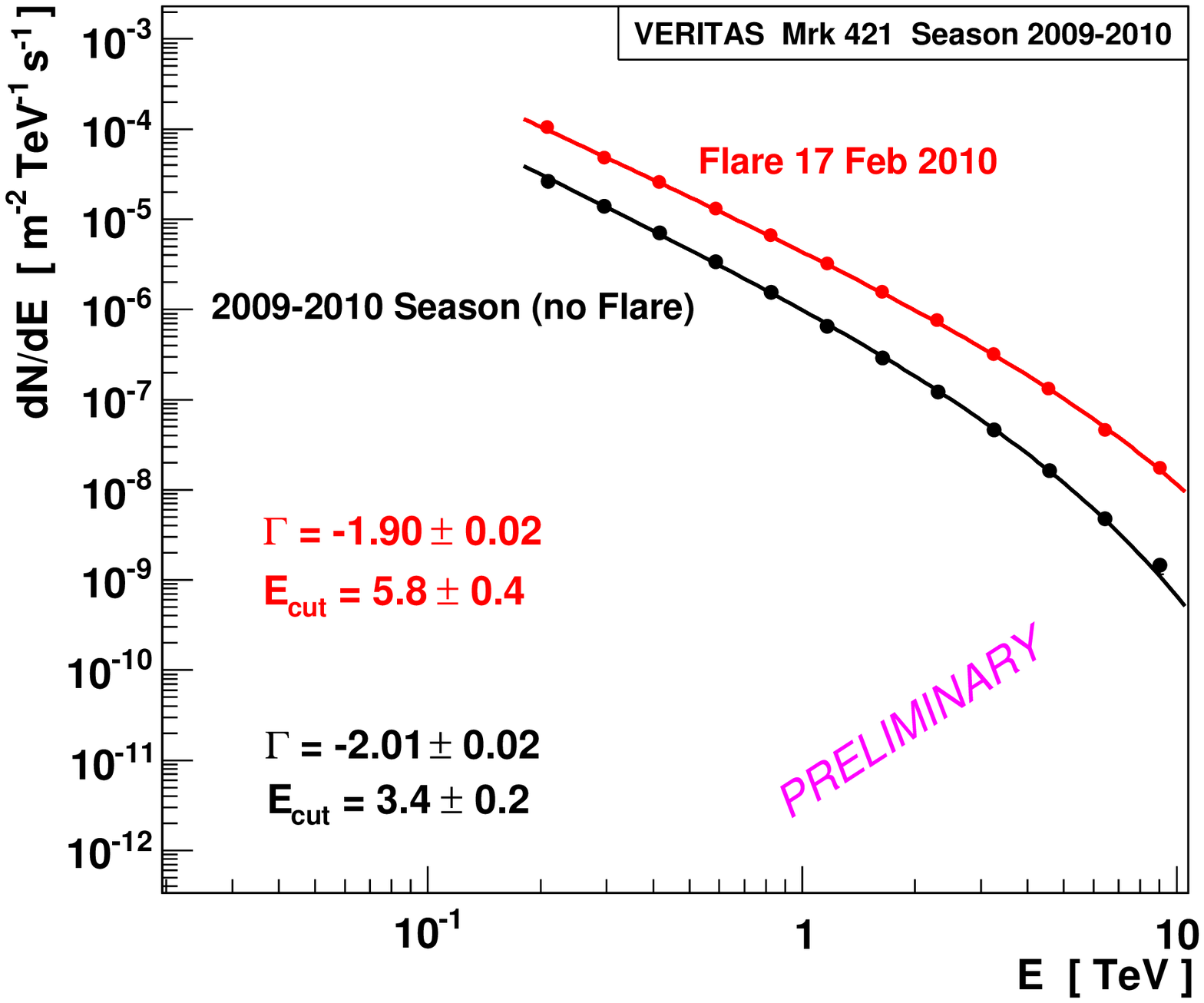}
  \caption{\emph{(top)}VERITAS $\theta^2$ plot of Markarian~421 data collected during the February 17, 2010
  flare. The shaded area is the background, the solid line represents the on-source events. VERITAS
  detects the HBL Markarian~421 at the $260\sigma$ significance level in 4.9~hr effective observation time,
  corresponding to approximatively 8 Crab unit flux. \emph{(bottom)} VERITAS preliminary spectral analysis of Markarian~421 for two different flux levels: the flare on February 17, 2010 (red dots) and the rest of the 2009-2010 season (black dots).}
  \label{fig1}
 \end{figure}

Top of figure~\ref{fig1} shows the $\theta^2$ plot
of the VERITAS data collected the night of the flare.
Although in partial-array configuration, VERITAS detected VHE emission from Markarian~421
at the $260\sigma$ significance level over 4.9~hr effective observation time.
The VHE photon rate is 1.2~Hz, corresponding to $\sim8$ times that of the Crab Nebula.
The average flux level during the flaring event is about four times the average flux
level during the monitoring campaign. A preliminary spectral analysis is done separately
for the night of the flare and for the rest of the season. Clear spectral evolution dependent on
the flux level is visible, consisting in spectral hardening and increase of the cutoff energy
in the high-flux state data.

\begin{figure*}[th]
  \centering
  \includegraphics[width=6.5in,height=3in]{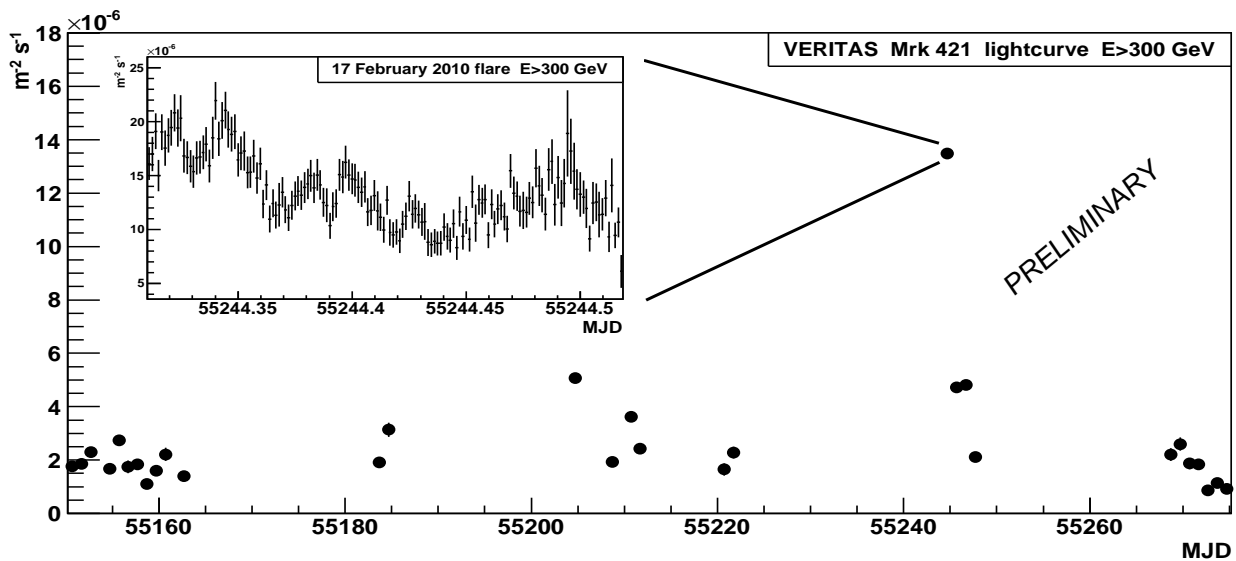}
  \caption{Nigthly lightcurve of Markarian~421 during the VERITAS 2009-2010 monitoring campaign.
  The source is detected at an average flux level of approximatively 2 Crab units over the entire season.
  On  February 17, 2010 the source is observed in flaring state at a flux level of $\sim8$ Crab units.
  A zoom of the intra-night 2-minute bins lightcurve of the flaring event is shown. 
  During the flaring event the source is detected at $>10\sigma$ significance in any of the 2-minute bin,
and intra-night variability is seen. }
  \label{fig2}
 \end{figure*}

Figure~\ref{fig2} shows the Markarian~421 nightly
lightcurve over the entire season, and for the February 17, 2010
flaring event.
During the flaring episode, the source is detected at $>10\sigma$ statistical
significance in any 2 minute bin. A preliminary time-resolved analysis
shows significant flux variability on a time scale of $\sim$5-10 minutes. Intra-night variability
in the May 2008 outburst from Markarian~421 is observed too,
although on a longer time scale~\cite{Matthias}.

\section*{Conclusions}

VERITAS is monitoring the brightest blazars in order to promptly observe and study the physics at work
in ultra-relativistic jets. The entire observational campaign on Markarian~421 between 2006 and 2006 provided a
detailed characterization of the broadband SED, supporting the one-zone leptonic SSC model
as the mechanism responsible for the observed non-thermal continuum emission.
The monitoring strategy is successful and resulted in coordinated MWL
observational campaigns  with optical/UV, X-ray and $\gamma$-ray partners.
Surprisingly, no obvious correlation between the lower energy (optical/UV and X-ray) and the
higher energy (VHE) components is seen in the 2006 and 2008 coordinated MWL campaigns.
More recently VERITAS observed a
big flare on February 17, 2010. The detection of a $\sim8$
Crab unit flux from Markarian~421 allows to make time-resolved analysis in order to study intra-night variability,
and place constraints on the size and energetics of the physical region responsible for the $\gamma$-ray 
emission. Such a temporal analysis and time-resolved spectral analysis are in progress. A correlated
study of the VERITAS VHE observation with the MWL observations provided by LAT and XRT,
which will provide insights on the parameterization of the physical models, is still in progress.

This research is supported by grants from the US Department of Energy, the US National Science Foundation, 
and the Smithsonian Institution, by NSERC in Canada, by Science Foundation Ireland, and by STFC in the UK. 
We acknowledge the excellent work of the technical support staff at the FLWO and at the collaborating 
institutions in the construction and operation of the instrument.


\clearpage

\end{document}